\documentclass[a4,12pt]{article}
\newcommand{\nc}{\newcommand}
\nc{\bg}{B. Grzadkowski}
\nc{\non}{\nonumber}
\def\dps{\displaystyle}
\def\mib#1{\mbox{\boldmath $#1$}}

\def\bra#1{\langle #1 |} \def\ket#1{|#1\rangle}
\def\vev#1{\langle #1\rangle}

\nc{\barx}{\bar{x}}\nc{\pbarn}{\;\hbox {pb}}\nc{\fbarn}{\;\hbox {fb}}
\nc{\hc}{\hbox {h.c.}} \nc{\re}{\hbox {Re}} 
\nc{\mev}{\hbox {MeV}} \nc{\gev}{\;\hbox {GeV}}
\def\gesim{\lower0.5ex\hbox{$\:\buildrel >\over\sim\:$}}
\def\lesim{\lower0.5ex\hbox{$\:\buildrel <\over\sim\:$}}
\nc{\prd}[3]{{\it Phys.\ Rev.}\ {{\bf D{#1}} (#2) #3}}
\nc{\prl}[3]{{\it Phys.\ Rev.\ Lett.}\ {{\bf {#1}} (#2), #3}}
\nc{\plb}[3]{{\it Phys.\ Lett.}\ {{\bf B{#1}} (#2) #3}}
\nc{\npb}[3]{{\it Nucl.\ Phys.}\ {{\bf B{#1}} (#2) #3}}
\nc{\ptp}[3]{{\it Prog.\ Theor.\ Phys.}\ {{\bf {#1}} (#2), #3}}
\nc{\zfp}[3]{{\it Z.\ Phys.}\ {{\bf C{#1}} (#2) #3}}
\nc{\epj}[3]{{\it Eur.\ Phys.\ J.}\ {{\bf C{#1}} (#2) #3}}
\nc{\mpla}[3]{{\it Mod.\ Phys.\ Lett.}\ {{\bf A{#1}} (#2) #3}}
\nc{\rmp}[3]{{\it Rev.\ Mod.\ Phys.}\ {{\bf {#1}} (#2) #3}}
\nc{\ijmpa}[3]{{\it Int.\ J.\ Mod.\ Phys.}\
               {{\bf A{#1}} (#2) #3}}
\nc{\ttbar}{t\bar{t}}         \nc{\bbbar}{b\bar{b}}
\nc{\tanb}{\tan \beta}        \nc{\twbdec}{t\to W^+ b}
\nc{\tbwbdec}{\bar{t}\to W^- \bar{b}}
\nc{\epem}{e^+e^-}            \nc{\eett}{\epem \to \ttbar}
\nc{\sigeett}{\sigma_{e\bar{e}\to\ttbar}}
\nc{\wpwm}{W^+W^-}            \nc{\tbar}{\bar{t}}
\nc{\bbar}{\bar{b}}           \nc{\wpp}{W^+}
\nc{\mt}{m_t}    \nc{\mts}{m_t^2}   \nc{\mw}{m_W}    \nc{\mws}{m_W^2}
\nc{\mz}{m_Z}    \nc{\mzs}{m_Z^2}
\nc{\ttbardec}{\ttbar \to W^+W^-\bbbar}
\nc{\wwbb}{W^+W^-\bbbar}      \nc{\sm}{SM}
\nc{\cw}{\cos\theta_W}        \nc{\sw}{\sin\theta_W}
\nc{\sws}{\sin^2\theta_W}     \nc{\sig}{\sigma_{tot}}
\nc{\lp}{{\ell}^+}              \nc{\lm}{{\ell}^-}
\nc{\epsl}{\epsilon_L}        \nc{\cp}{C\!P}
\nc{\gaga}{\gamma\gamma}
\nc{\splus}{s_+}       \nc{\smin}{s_-}        \nc{\eps}{\epsilon}
\nc{\psp}{Ps_+}        \nc{\psm}{Ps_-}        \nc{\lsp}{ls_+}
\nc{\lsm}{ls_-}        \nc{\sss}{s_+s_-}      \nc{\m}{m_t}
\nc{\mq}{m_t^2}        \nc{\mr}{\frac{1}{\m}} \nc{\av}{A_{\gamma}}
\nc{\bv}{B_{\gamma}}   \nc{\az}{A_Z}          \nc{\bz}{B_Z}
\nc{\avs}{A_{\gamma}^2}\nc{\azs}{A_Z^2}       \nc{\bzs}{B_Z^2}
\nc{\dav}{\delta \! A_{\gamma}}   \nc{\dbv}{\delta \! B_{\gamma}}
\nc{\dcv}{\delta C_{\gamma}}      \nc{\ddv}{\delta \! D_{\gamma}}
\nc{\daz}{\delta \! A_Z}          \nc{\dbz}{\delta \! B_Z}
\nc{\dcz}{\delta C_Z}             \nc{\ddz}{\delta \! D_Z}
\nc{\dev}{\delta \! E_{\gamma}}   \nc{\dez}{\delta \! E_Z}
\nc{\dfv}{\delta \! F_{\gamma}}   \nc{\dfz}{\delta \! F_Z}
\nc{\rdav}{{\rm Re}(\delta \! A_{\gamma}) \:}
\nc{\rdbv}{{\rm Re}(\delta \! B_{\gamma}) \:}
\nc{\rdcv}{{\rm Re}(\delta C_{\gamma}) \:}
\nc{\rddv}{{\rm Re}(\delta \! D_{\gamma}) \:}
\nc{\rdaz}{{\rm Re}(\delta \! A_Z) \:}
\nc{\rdbz}{{\rm Re}(\delta \! B_Z) \:}
\nc{\rdcz}{{\rm Re}(\delta C_Z) \:}
\nc{\rddz}{{\rm Re}(\delta \! D_Z) \:}
\nc{\idav}{{\rm Im}(\delta \! A_{\gamma}) \:}
\nc{\idbv}{{\rm Im}(\delta \! B_{\gamma}) \:}
\nc{\idcv}{{\rm Im}(\delta C_{\gamma}) \:}
\nc{\iddv}{{\rm Im}(\delta \! D_{\gamma}) \:}
\nc{\idaz}{{\rm Im}(\delta \! A_Z) \:}
\nc{\idbz}{{\rm Im}(\delta \! B_Z) \:}
\nc{\idcz}{{\rm Im}(\delta C_Z) \:}
\nc{\iddz}{{\rm Im}(\delta \! D_Z) \:}
\nc{\cz}{(1+v_e^2)d\:\!'^2}         \nc{\ci}{v_ed\:\!'}
\nc{\ccz}{v_ed\:\!'^2}              \nc{\cci}{d\:\!'}
\nc{\lspace}{\;\;\;\;\;\;\;\;\;\;}  \nc{\llspace}{\lspace \lspace}
\nc{\beq}{\begin{equation}}   \nc{\eeq}{\end{equation}}
\nc{\bea}{\begin{eqnarray}}   \nc{\eea}{\end{eqnarray}}
\nc{\baa}{\begin{array}}      \nc{\eaa}{\end{array}}
\nc{\bit}{\begin{itemize}}    \nc{\eit}{\end{itemize}}
\nc{\ben}{\begin{enumerate}}  \nc{\een}{\end{enumerate}}
\nc{\bce}{\begin{center}}     \nc{\ece}{\end{center}}
\nc{\ocal}{{\cal O}}
\newcounter{QQ}
\newcounter{QQF}
\newcounter{QQE}
\begin{document}
\pagestyle{empty} \setlength{\footskip}{2.0cm}
\setlength{\oddsidemargin}{0.5cm} \setlength{\evensidemargin}{0.5cm}
\renewcommand{\thepage}{-- \arabic{page} --}
\def\mib#1{\mbox{\boldmath $#1$}}
\def\bra#1{\langle #1 |}      \def\ket#1{|#1\rangle}
\def\vev#1{\langle #1\rangle} \def\dps{\displaystyle}
\nc{\tb}{\stackrel{{\scriptscriptstyle (-)}}{t}}
\nc{\bb}{\stackrel{{\scriptscriptstyle (-)}}{b}}
\nc{\fb}{\stackrel{{\scriptscriptstyle (-)}}{f}}
\nc{\pp}{\gamma \gamma}
\nc{\pptt}{\pp \to \ttbar}
   \def\thebibliography#1{\centerline{REFERENCES}
     \list{[\arabic{enumi}]}{\settowidth\labelwidth{[#1]}\leftmargin
     \labelwidth\advance\leftmargin\labelsep\usecounter{enumi}}
     \def\newblock{\hskip .11em plus .33em minus -.07em}\sloppy
     \clubpenalty4000\widowpenalty4000\sfcode`\.=1000\relax}\let
     \endthebibliography=\endlist
   \def\sec#1{\addtocounter{section}{1}\section*{\hspace*{-0.72cm}
     \normalsize\bf\arabic{section}.$\;$#1}\vspace*{-0.3cm}}
\vspace*{-1.7cm}
\noindent

 \vspace{-0.7cm}
\begin{flushright}
$\vcenter{
\hbox{{\footnotesize IFT-28-05~~~~FUT-05-02}}
\hbox{{\footnotesize UCRHEP-T399}}
\hbox{{\footnotesize TOKUSHIMA Report}}
\hbox{(hep-ph/0511038)}
}$
\end{flushright}
\renewcommand{\thefootnote}{$\dag$}
\vskip 0.6cm
\begin{center}
{\large\bf Studying Possible Anomalous Top-Quark Couplings}

\vskip 0.15cm
{\large\bf at Photon Colliders}\footnote{Presented by K. Ohkuma at the 29th International Conference
of Theoretical Physics
''Matter To The Deepest: Recent Developments in Physics of
Fundamental Interactions'',
Ustron (Poland), September 8--14, 2005.\\
E-mail address: \tt ohkuma@fukui-ut.ac.jp}
\end{center}

\vspace{0.2cm}
\begin{center}
\renewcommand{\thefootnote}{\alph{footnote})}
{\sc Bohdan GRZADKOWSKI$^{\:1) \:}$}, \  
{\sc Zenr\=o HIOKI$^{\:2) \:}$}

\vskip 0.15cm
{\sc Kazumasa OHKUMA$^{\:3) \:}$}\ 
\ and \ \ 
{\sc Jos\'e WUDKA$^{\:4) \:}$}
\end{center}

\vspace*{0.2cm}
\centerline{\sl $1)$ Institute of Theoretical Physics,\ Warsaw
University}
\centerline{\sl Ho\.za 69, PL-00-681 Warsaw, Poland}

\vskip 0.2cm
\centerline{\sl $2)$ Institute of Theoretical Physics,\
University of Tokushima}
\centerline{\sl Tokushima 770-8502, Japan}

\vskip 0.2cm
\centerline{\sl $3)$ Department of Management Science,\
Fukui University of Technology}
\centerline{\sl Fukui 910-8505, Japan}

\vskip 0.2cm
\centerline{\sl $4)$ Department of Physics,\
University of California}
\centerline{\sl Riverside, CA 92521-0413, USA}

\vspace*{0.9cm}
\centerline{ABSTRACT}

\vspace*{0.3cm}
\baselineskip=20pt plus 0.1pt minus 0.1pt
Search for new-physics through possible anomalous $t\bar{t}\gamma$, 
$tbW$ and $\gamma\gamma H$ couplings which are generated by
$SU(2)\times U(1)$ gauge-in\-var\-i\-ant dimension-6 effective
operators is discussed, using energy and angular distributions
of final charged-lepton/$b$-quark in $\gamma\gamma\to t\tbar
\to \ell X/bX$ for various beam polarizations. 
Optimal beam polarizations that 
minimize uncertainty in determination 
of those non-standard couplings are found
performing an optimal-observable analysis.
\vspace*{0.4cm} \vfill

PACS:  14.65.Fy, 14.65.Ha, 14.70.Bh

Keywords:
anomalous top-quark couplings, $\gamma\gamma$ colliders \\

\newpage
\renewcommand{\thefootnote}{$\sharp$\arabic{footnote}}
\pagestyle{plain} \setcounter{footnote}{0}
\pagestyle{plain} \setcounter{page}{1}
\baselineskip=21.0pt plus 0.2pt minus 0.1pt

\sec{Introduction}
Although more than ten years have passed since the discovery
of the top quark at Fermilab Tevatron \cite{Abe:1994xt},
this collider is still the only facility which can produce
the top quark and  top properties have not been well determined yet.
In the near future, however, a more powerful top-quark factory
will be realized at Large Hadron Collider (LHC)
\cite{Beaudette:2005nj} and/or International Linear Collider
(ILC) \cite{LP2005}. It is therefore definitely meaningful to
get prepared for performing analyses assuming substantial
top-quark data.

Since the top-quark
mass, $m_t$, is of the order of the electroweak scale, it is
quite reasonable to hope that measurements of top couplings
and width could reveal some features of physics beyond the
Standard Model (SM). The huge $m_t$ also provides us some
practical advantages, e.g., the top quark decays before it
hadronizes and therefore experimental data are going to be
free from any substantial contamination by unknown bound
state effects~\cite{Bigi:1980az}. Consequently, one
can easily get information concerning top-quark couplings
via distributions of its decay products~\cite{Kuhn:1983ix}.
Furthermore, since the Yukawa coupling of the top quark is much
larger than that of other particles observed to date, the
top quark must be very sensitive to Higgs
boson. Thus, the top quark could also be useful while testing
extensions of the scalar sector of the SM.

Motivated by the above comments, we have carried out an 
analysis~\cite{Grzadkowski:2003tf,Grzadkowski:2005ye} of
top-quark production and decay at photon colliders
\cite{Ginzburg:1981vm,Borden:1992qd}. We have considered
the charged-lepton/$b$-quark momentum distributions in the
process $\gamma \gamma \to t\bar{t} \to \ell X / b X$,
focusing on possible signals of new physics. Here we are
going to present main findings that we obtained up to date:
after describing our basic framework in sec. 2, we show
results of our optimal analysis \cite{optimal} in sec. 3.
A brief summary and discussion are contained in sec. 4.

\sec{Framework}
In order to describe possible new-physics effects,
we have used an effective low-energy Lagrangian 
\cite{Buchmuller:1986jz}, i.e., the SM Lagrangian is modified by
the addition of a series of $SU(3)\times SU(2)\times U(1)$
gauge-invariant operators, which are suppressed by inverse
powers of a new-physics scale ${\mit\Lambda}$. Among those
operators, the largest contribution comes from dimension-6
operators,\footnote{Dimension-5 operators are not included
    since they violate lepton number \cite{Buchmuller:1986jz}
    and are irrelevant for the processes considered here.}
denoted as ${\cal O}_i$, and we have the effective Lagrangian as
\begin{equation}
{\cal L}_{\rm eff}={\cal L}_{\rm SM}
+\frac1{{\mit\Lambda}^2}\sum_i (\alpha_i {\cal O}_i
+ {\rm H.c.}) + O( {\mit\Lambda}^{-3} ).
\end{equation}
The operators
relevant here (for more details see \cite{Grzadkowski:2003tf,
Grzadkowski:2005ye})
lead to the following non-standard top-quark- and
Higgs-boson-couplings:
(1) $C\!P$-conserving $t\bar{t}\gamma$ vertex,
(2) $C\!P$-violating $t\bar{t}\gamma$ vertex,
(3) $C\!P$-conserving $\gamma\gamma H$ vertex,
(4) $C\!P$-violating $\gamma\gamma H$ vertex, and
(5) anomalous $tbW$ vertex. We expressed the size of their
strength in terms of five independent parameters
$\alpha_{\gamma 1}$, $\alpha_{\gamma 2}$, $\alpha_{h1}$,
$\alpha_{h2}$ and $\alpha_{d}$. The explicit forms of
these anomalous couplings in terms of the coefficients
of dimension-6 operators are to be found in
\cite{Grzadkowski:2003tf,Grzadkowski:2005ye}.

The initial-state polarizations are characterized by the
initial electron and positron longitudinal polarizations $P_e$
and $P_{\bar{e}}$, the maximum average linear polarizations
$P_t$ and $P_{\tilde{t}}$ of the initial-laser photons with
the azimuthal angles $\varphi$ and $\tilde{\varphi}$
(defined in the same way as in \cite{Ginzburg:1981vm}), and
their average helicities $P_{\gamma}$ and $P_{\tilde{\gamma}}$.
The photonic polarizations $P_{t,\gamma}$ and
$P_{\tilde{t},\tilde{\gamma}}$ have to satisfy
\begin{equation}
0 \leq P_t^2 + P_{\gamma}^2 \leq 1,
\ \ \ \ \ \ \ \ \
0 \leq P_{\tilde{t}}^2 + P_{\tilde{\gamma}}^2 \leq 1.
\end{equation}
For the linear polarization, we denote the relative azimuthal
angle by $\chi \equiv \varphi-\tilde{\varphi}$. In order to
find its optimal value, we studied the $\chi$-dependence of
$\sigma(\gamma\gamma\to t\bar{t})$ including
$\alpha_{\gamma 1,\gamma2,h1,h2}$ terms. As a result, we
found the $\alpha_{\gamma2}$ and $\alpha_{h2}$ terms are
sensitive to $\chi$ with the maximal sensitivity at
$\chi=\pi/4$ as long as we are not too close to the Higgs-pole,
while the others did not lead to any relevant dependence.
This has also been noticed in \cite{Choi:1995kp} concerning
the $\alpha_{\gamma 2}$ term. Therefore we fix $\chi$ to be
$\pi/4$.

In deriving distributions of secondary fermions $(=\ell/b)$
we have treated the decaying $t$ and $W$ as on-shell particles.
We have also neglected contributions that are quadratic in
$\alpha_i$ ($i=\gamma 1,\gamma 2, h1,h2,d$). Therefore the
energy-angular distributions of $\ell/b$ in the $e\bar{e}$
CM frame\footnote{Following the standard approach
    \cite{Ginzburg:1981vm}, each photonic beam originates
    as a laser beam back-scattered on electron ($e$) or
    positron ($\bar{e}$) beam. Therefore the $e\bar{e}$ CM
    frame refers to those initial electron-positron beams.}
can be expressed as
\begin{equation}
\frac{d\sigma}{dE_{\ell/b} d\cos\theta_{\ell/b}}
=f_{\rm SM}(E_{\ell/b}, \cos\theta_{\ell/b})
 + \sum_i \alpha_i f_i (E_{\ell/b}, \cos\theta_{\ell/b}),
\label{distribution}
\end{equation}
where $f_{\rm SM}$ and $f_i$ are calculable functions:
$f_{\rm SM}$ denotes the SM contribution,
$f_{\gamma 1,\gamma 2}$ describe the anomalous
$C\!P$-conserving and $C\!P$-violating
$t\bar{t}\gamma$-ver\-ti\-ces contributions respectively,
$f_{h1,h2}$  those generated by the anomalous $C\!P$-conserving
and $C\!P$-violating $\gamma\gamma H$-ver\-ti\-ces,
and $f_d$ that by the anomalous $tbW$-vertex.

In order to apply the Optimal Observable (OO) method (see \cite{optimal} 
for details) to eq.(\ref{distribution}), we first have to
calculate the following matrix elements using $f_{\rm SM}$
and $f_i$
\begin{eqnarray}
&&{\cal M}_{ij}=\int dE_{\ell/b} d\cos\theta_{\ell/b}\nonumber\\
&&\ \ \ \ \ \ \ \ \ \ \ \ \times
f_i(E_{\ell/b}, \cos\theta_{\ell/b})
f_j(E_{\ell/b}, \cos\theta_{\ell/b})/
f_{\rm SM}(E_{\ell/b}, \cos\theta_{\ell/b})~~~~
\end{eqnarray}
and its inverse matrix $X_{ij}$, where $i,j=1,\cdots, 6$
correspond to SM, $\gamma 1$, $\gamma 2$, $h1$, $h2$ and $d$
respectively. Then, according to \cite{optimal}, the expected
statistical uncertainty for the measurements of $\alpha_i$
is given by
\beq
{\mit\Delta}\alpha_i=\sqrt{I_0 X_{ii}/N_{\ell/b}},
\eeq
where
\[
I_0\equiv \int dE_{\ell/b}
d\cos\theta_{\ell/b}\, f_{\rm SM}(E_{\ell/b}, \cos\theta_{\ell/b})
\]
and $N_{\ell/b}$ is the total number of collected events.
Since we are not stepping into the Higgs-resonance region,
we simply compute $N_{\ell/b}$ from the SM total cross section
multiplied by the lepton/$b$-quark detection efficiency
$\epsilon_{\ell/b}$ and the integrated $e\bar{e}$ luminosity
$L_{e\bar{e}}$, which leads to $N_{\ell/b}$ independent of
$m_H$.

Concerning the effective Lagrangian approach, readers might
wonder why we did not follow the same strategy as in
$e\bar{e}\to t\bar{t}\to \ell X/bX$ analysis, where we started
from the most general invariant amplitude with non-local
(i.e., in general momentum-dependent) form factors
\cite{Grzadkowski:1996kn,Grzadkowski:2000nx}. As a matter of fact,
such an approach is not possible for $\gamma\gamma\to t\bar{t}$
because of the virtual top-quark line appearing in the $t$-channel
amplitudes. In case of $e\bar{e}\to t\bar{t}$, all the kinematical
variables on which the form factors may depend are fixed for a
given $\sqrt{s}$ and consequently we can treat all those form
factors as constants, while this is not the case for
$\gamma\gamma\to t\bar{t}$.~\footnote{For more details see the  
discussion in section 4 of~\cite{Grzadkowski:2005ye}} 

\sec{Numerical analysis and results}

In ref.\cite{Grzadkowski:2003tf}, where our main
concern was to construct a fundamental framework for practical
analysis, we used (1) $P_e=P_{\bar{e}}=1$ and
$P_t =P_{\tilde{t}}=P_{\gamma}=P_{\tilde{\gamma}}=1/\sqrt{2}$,
and (2) $P_e=P_{\bar{e}}=P_{\gamma}=P_{\tilde{\gamma}}=1$ as
typical polarization examples and performed an OO-analysis.
Inverting the matrix ${\cal M}_{ij}$, we have noticed that the
numerical results for $X_{ij}$ are often unstable
\cite{Grzadkowski:2003tf}: even a tiny variation of ${\cal M}_{ij}$
changes $X_{ij}$ significantly. This indicates that some of
$f_i$ have similar shapes and therefore their coefficients
cannot be disentangled easily.
The presence of such instability has forced us to refrain from
determining all the couplings at once through this process alone.
That is, we have assumed that some of $\alpha_i$'s had been
measured in other processes (e.g., in
$e\bar{e}\to t\bar{t}\to{\ell}^{\pm}X$), and we performed an
analysis with smaller number of independent parameters.

When estimating the statistical uncertainty in simultaneous
measurements, e.g., of $\alpha_{\gamma 1}$ and $\alpha_{h 1}$
(assuming all other coefficients are known), we need only
the components with indices 1, 2 and 4. In such a ``reduced
analysis'', we still encountered the instability problem, and
we selected ''stable solutions'' according to the following
criterion: Let us express the resultant uncertainties as
${\mit\Delta}\alpha_{\gamma 1}^{[3]}$ and
${\mit\Delta}\alpha_{h 1}^{[3]}$, where ``3'' shows that we
use the input ${\cal M}_{ij}$, keeping three decimal places.
In addition, we also compute
${\mit\Delta}\alpha_{\gamma 1}^{[2]}$
and ${\mit\Delta}\alpha_{h 1}^{[2]}$
by rounding ${\cal M}_{ij}$
off to two decimal places. Then, we accept the result as a
stable solution if both of the deviations
$|{\mit\Delta}\alpha_{\gamma 1,h 1}^{[3]}
-{\mit\Delta}\alpha_{\gamma 1,h 1}^{[2]}|/
{\mit\Delta}\alpha_{\gamma 1,h 1}^{[3]}$ are less than 10 \%.

In \cite{Grzadkowski:2005ye}, varying polarization parameters as
$P_{e,\bar{e}}=0,\,\pm1$, $P_{t,\tilde{t}}=0,
\,1/\sqrt{2},\,$ 1, and $P_{\gamma,\tilde{\gamma}}=0,
\,\pm 1/\sqrt{2},\,\pm1$, we searched for the combinations that
could make the statistical uncertainties ${\mit\Delta} \alpha_i$
minimum for $\sqrt{s_{e\bar{e}}}=500$ GeV and ${\mit\Lambda}=1$
TeV. We also changed the Higgs mass as $m_H=$100, 300 and 500 GeV,
which lead to the width ${\mit\Gamma}_H=1.08\times 10^{-2}$, 8.38
and 73.4 GeV respectively according to the standard-model formula.

Although we did not find again any stable solution in the four-
and five-parameter analysis, we did find some solutions not only
in the two- but also in the three-parameter analysis. This is
quite in contrast to the results in \cite{Grzadkowski:2003tf},
where we had no stable solution for the three-parameter analysis.
However, since not all the stable solutions gave us good
statistical precision, we adopted only those which satisfy
the following conditions:
\begin{itemize}
\item Three-parameter analysis \\
       At least two unknown couplings of three could be
       determined with accuracy better than 0.1
       for a integrated luminosity of
       $L_{e\bar{e}}=500\ {\rm fb}^{-1}$ without
       detection-efficiency suppression (i.e., $\epsilon_{\ell/b}=1$).
\item Two-parameter analysis \\
We found many stable solutions, therefore for illustration
we adopt the following strategy:
\bit
\item we choose a final state (charged lepton or bottom quark),
\item we fix the Higgs-boson mass $m_H$,
\item for each pair of ${\mit\Delta}\alpha_i$ and
 ${\mit\Delta}\alpha_j$ that satisfy
 ${\mit\Delta}\alpha_{i,j}\leq 0.1$ for the luminosity of
 $L_{e\bar{e}}=500\ {\rm fb}^{-1}$ we show only those that make
$({\mit\Delta}\alpha_i)^2+({\mit\Delta}\alpha_j)^2$ minimum.
\eit
\end{itemize}

The results are presented below. We did not fix the detection
efficiencies $\epsilon_{\ell/b}$ since they depend
on detector parameters and will get better with development of
detection technology.

\setcounter{QQ}{\arabic{equation}} \addtocounter{QQ}{1}
\begin{description}
\item[1)] Three parameter analysis
\item[] $\oplus$ Final charged-lepton detection
\begin{description}
  \item[] $m_{H}=500$ GeV
  \item[$\bullet$]
    $P_e=P_{\bar{e}}=0,~P_t = P_{\tilde{t}}=1/\sqrt{2},~
     P_\gamma = - P_{\tilde{\gamma}}=1/\sqrt{2}$,
     $N_{\ell}\simeq 6.1\times 10^3 \epsilon_{\ell}$\\
     ${\mit\Delta} \alpha_{\gamma2}=0.94/\sqrt{\epsilon_{\ell}},~
      {\mit\Delta} \alpha_{h2}=0.11/\sqrt{\epsilon_{\ell}} ,~
      {\mit\Delta} \alpha_{d}=0.042/\sqrt{\epsilon_{\ell}}$.
\hfill (\arabic{QQ})

Strictly speaking, this result does not satisfy our condition
for the three-parameter analysis, but we show it since
${\mit\Delta} \alpha_{h2}$ exceeds the limit by only 0.01. 
\setcounter{QQF}{\arabic{QQ}} \addtocounter{QQ}{1}
\end{description}
\item[] $\oplus$ Final bottom-quark detection
\begin{description}
  \item[] $m_{H}=100$ GeV
  \item[$\bullet$]
    $P_e=P_{\bar{e}}=1,~P_t = P_{\tilde{t}}=1/\sqrt{2},~
     P_\gamma = - P_{\tilde{\gamma}}=1/\sqrt{2}$,
$N_{b}\simeq 4.2\times 10^4 \epsilon_{b}$\\
     ${\mit\Delta} \alpha_{h1}=0.086/\sqrt{\epsilon_{b}}$,~
     ${\mit\Delta} \alpha_{h2}=0.21/\sqrt{\epsilon_{b}}$,~
     ${\mit\Delta} \alpha_{d}=0.037/\sqrt{\epsilon_{b}}$.
\hfill (\arabic{QQ})\addtocounter{QQ}{1}
  \item[] $m_{H}=500$ GeV
  \item[$\bullet$]
    $P_e=P_{\bar{e}}=0,~P_t = P_{\tilde{t}}=1/\sqrt{2},~
     P_\gamma = - P_{\tilde{\gamma}}=1/\sqrt{2}$,
     $N_{b}\simeq 2.8\times 10^4 \epsilon_{b}$\\
     ${\mit\Delta} \alpha_{\gamma2}=0.61/\sqrt{\epsilon_{b}}$,~
     ${\mit\Delta} \alpha_{h2}=0.054/\sqrt{\epsilon_{b}}$,~
     ${\mit\Delta} \alpha_{d}=0.052/\sqrt{\epsilon_{b}}$.
\hfill (\arabic{QQ})\addtocounter{QQ}{1}
 \end{description}
\item[2)] Two parameter analysis
\item[] $\oplus$ Final charged-lepton detection
\begin{description}
    \item[] Independent of $m_H$
%
  \item[$\bullet$]
    $P_e=P_{\bar{e}}=-1,~P_t = P_{\tilde{t}}=1,~
     P_\gamma = P_{\tilde{\gamma}}=0$,
     $N_{\ell}\simeq 1.0\times 10^4 \epsilon_{\ell}$\\
     ${\mit\Delta} \alpha_{\gamma1}=0.051/\sqrt{\epsilon_{\ell}}$,~
     ${\mit\Delta} \alpha_{d}=0.022/\sqrt{\epsilon_{\ell}}$.
\hfill (\arabic{QQ})\addtocounter{QQ}{1}

This result is free from $m_H$ dependence since the Higgs-exchange
diagrams do not contribute to $\alpha_{\gamma1}$ and $\alpha_d$
determination within our approximation.
  \item[] $m_H=100$ GeV
  \item[$\bullet$]
    $P_e=P_{\bar{e}}=-1,~P_t = P_{\tilde{t}}=1/\sqrt{2},~
     P_\gamma = P_{\tilde{\gamma}}=1/\sqrt{2}$,
     $N_{\ell}\simeq 1.9\times 10^4 \epsilon_{\ell}$\\
     ${\mit\Delta} \alpha_{h1}=0.034/\sqrt{\epsilon_{\ell}}$,~
     ${\mit\Delta} \alpha_{d}=0.017/\sqrt{\epsilon_{\ell}}$.
\hfill (\arabic{QQ})\addtocounter{QQ}{1}
  \item[] $m_H=300$ GeV
  \item[$\bullet$]
    $P_e=P_{\bar{e}}=-1,~P_t = P_{\tilde{t}}=0,~
     P_\gamma = P_{\tilde{\gamma}}=1$, 
     $N_{\ell}\simeq 2.4\times 10^4 \epsilon_{\ell}$\\
     ${\mit\Delta} \alpha_{h1}=0.013/\sqrt{\epsilon_{\ell}}$,~
     ${\mit\Delta} \alpha_{d}=0.015/\sqrt{\epsilon_{\ell}}$.
\hfill (\arabic{QQ})\addtocounter{QQ}{1}
  \item[] $m_H=500$ GeV
  \item[$\bullet$]
    $P_e=P_{\bar{e}}=-1,~P_t = P_{\tilde{t}}=0,~
     P_\gamma = P_{\tilde{\gamma}}=1$, 
     $N_{\ell}\simeq 2.4\times 10^4 \epsilon_{\ell}$\\
     ${\mit\Delta} \alpha_{h1}=0.023/\sqrt{\epsilon_{\ell}}$,~
     ${\mit\Delta} \alpha_{d}=0.015/\sqrt{\epsilon_{\ell}}$.
\hfill (\arabic{QQ})\addtocounter{QQ}{1}
\item[$\bullet$]
    $P_e=P_{\bar{e}}=-1,~P_t = P_{\tilde{t}}=0,~
     P_\gamma = P_{\tilde{\gamma}}=1$, 
     $N_{\ell}\simeq 2.4\times 10^4 \epsilon_{\ell}$\\
     ${\mit\Delta} \alpha_{h2}=0.030/\sqrt{\epsilon_{\ell}}$,~
     ${\mit\Delta} \alpha_{d}=0.015/\sqrt{\epsilon_{\ell}}$.
\hfill (\arabic{QQ})\addtocounter{QQ}{1}
\end{description}
\item[] $\oplus$ Final bottom-quark detection
\begin{description}
  \item[] $m_H=100$ GeV
  \item[$\bullet$]
    $P_e=P_{\bar{e}}=-1,~P_t = P_{\tilde{t}}=1/\sqrt{2},~
     P_\gamma = - P_{\tilde{\gamma}}=-1/\sqrt{2}$,\\ 
     \hspace*{-0.5cm}$N_{b}\simeq 4.2\times 10^4 \epsilon_{b}$\\
     ${\mit\Delta} \alpha_{h1}=0.058/\sqrt{\epsilon_{b}}$,~
     ${\mit\Delta} \alpha_{d}=0.026/\sqrt{\epsilon_{b}}$.
\hfill (\arabic{QQ})\addtocounter{QQ}{1}
  \item[] $m_H=300$ GeV
  \item[$\bullet$]
    $P_e=P_{\bar{e}}=-1,~P_t = P_{\tilde{t}}=1/\sqrt{2},~
     P_\gamma = -P_{\tilde{\gamma}}=-1/\sqrt{2},\\ 
   \hspace*{-0.5cm}  N_{b}\simeq 4.2\times 10^4 \epsilon_{b}$\\
     ${\mit\Delta} \alpha_{h1}=0.009/\sqrt{\epsilon_{b}}$,~
     ${\mit\Delta} \alpha_{h2}=0.074/\sqrt{\epsilon_{b}}$.
\hfill (\arabic{QQ})\addtocounter{QQ}{1}
  \item[$\bullet$]
    $P_e=P_{\bar{e}}=1,~P_t = P_{\tilde{t}}=1/\sqrt{2},~
     P_\gamma = -P_{\tilde{\gamma}}=-1/\sqrt{2}$, \\
    \hspace*{-0.6cm} $N_{b}\simeq 4.2\times 10^4 \epsilon_{b}$\\
     ${\mit\Delta} \alpha_{h1}=0.025/\sqrt{\epsilon_{b}}$,~
     ${\mit\Delta} \alpha_{d}=0.019/\sqrt{\epsilon_{b}}$.
\hfill (\arabic{QQ})\addtocounter{QQ}{1}
 \item[$\bullet$]
    $P_e=P_{\bar{e}}=1,~P_t = P_{\tilde{t}}=1/\sqrt{2},~
     P_\gamma = -P_{\tilde{\gamma}}=1/\sqrt{2}$, 
     $N_{b}\simeq 4.2\times 10^4 \epsilon_{b}$\\
     ${\mit\Delta} \alpha_{h2}=0.065/\sqrt{\epsilon_{b}}$,~
     ${\mit\Delta} \alpha_{d}=0.010/\sqrt{\epsilon_{b}}$.
\hfill (\arabic{QQ})\addtocounter{QQ}{1}
  \item[] $m_H=500$ GeV
  \item[$\bullet$]
    $P_e=P_{\bar{e}}=-1,~P_t = P_{\tilde{t}}=1,~
     P_\gamma = P_{\tilde{\gamma}}=0$, 
     $N_{b}\simeq 4.6\times 10^4 \epsilon_{b}$\\
     ${\mit\Delta} \alpha_{h1}=0.030/\sqrt{\epsilon_{b}}$,~
     ${\mit\Delta} \alpha_{d}=0.018/\sqrt{\epsilon_{b}}$.
\hfill (\arabic{QQ})\addtocounter{QQ}{1}
 \item[$\bullet$]
    $P_e=P_{\bar{e}}=-1,~P_t = P_{\tilde{t}}=1,~
     P_\gamma = P_{\tilde{\gamma}}=0$, 
     $N_{b}\simeq 4.6\times 10^4 \epsilon_{b}$\\
     ${\mit\Delta} \alpha_{h2}=0.028/\sqrt{\epsilon_{b}}$,~
     ${\mit\Delta} \alpha_{d}=0.014/\sqrt{\epsilon_{b}}$. 
\hfill (\arabic{QQ})\setcounter{QQE}{\arabic{QQ}}
\end{description}
\end{description}
\setcounter{equation}{\arabic{QQ}}
Using these results one can find (for known $m_H$) the most
suitable polarization for a determination of a given pair of
coefficients.

Note that it is difficult to determine $\alpha_{\gamma 1}$
and $\alpha_{\gamma 2}$ together for two- and three-parameter
analysis. Although we have found some
stable solutions that would allow for a determination of
$\alpha_{\gamma 1}$ in the lepton analysis, which we did not
find in \cite{Grzadkowski:2003tf}, the expected
precision is rather low. Nevertheless this is telling us that
the use of purely linear polarization for the laser is crucial
for measuring $\alpha_{\gamma 1}$. Unfortunately,
the statistical uncertainty of $\alpha_{\gamma 2}$ is still
large even in this analysis, so we did not list it as
solutions which gave us good statistical precisions. Therefore,
we have to look for other suitable processes to determine this
parameter, for a review see \cite{Atwood:2000tu}.

It was found that there are many combinations of polarization
parameters that make uncertainties of $\alpha_{h1,h2}$ and
$\alpha_{d}$ relatively small. For instance, analyzing the
$b$-quark final state with the polarization
    $P_e=P_{\bar{e}}=-1$, $P_t = P_{\tilde{t}}=1/\sqrt{2}$,
    $P_\gamma = -P_{\tilde{\gamma}}=-1/\sqrt{2}$
enables us to probe the properties of Higgs bosons whose mass
is around 300~GeV through the determination of $\alpha_{h1}$
and $\alpha_{h2}$.

As mentioned, the results are obtained for
${\mit\Lambda}=$ 1 TeV. If one assumes the new-physics scale to
be ${\mit\Lambda}=\lambda$ TeV, then all the above
results (${\mit\Delta}\alpha_i$) are replaced with
${\mit\Delta}\alpha_i/\lambda^2$, which means that the right-hand
sides of eqs.(\arabic{QQF})--(\arabic{QQE}) giving
${\mit\Delta}\alpha_i$ are all multiplied by $\lambda^2$.

Some additional comments are here in order.
\begin{itemize}
\item 
If we were going
to measure just the decay coefficient $\alpha_d$, then the optimal
polarization would be simply such that makes the top-production
rate largest with no Higgs exchange (this is because we keep
only linear terms in the anomalous couplings).
However, if
$\alpha_d$ and $\alpha_{h1}$ or $\alpha_{h2}$ are to be determined
then certain compromise of the SM $t\bar{t}$ production rate is
necessary as one needs the Higgs-boson exchange diagram as well.
\item 
If, on the other hand, only Higgs couplings are to be measured,
then the optimal polarization would make the Higgs-exchange
diagram as large as possible. It is obvious that for the most
precise determination of the $\gaga H$ couplings, one should
go to the resonance region in order to increase the Higgs
production rate. A detailed study of $C\!P$ violating effects
in $\gaga \to H$ has been performed, e.g., in \cite{Gounaris:1997ef}.
There, for the luminosity $L_{e\bar e} = 20$~fb$^{-1}$, the authors
estimate 3-$\sigma$ limits for $\alpha_{h2}$
($d_{\gaga}=(v/{\mit\Lambda})^2\alpha_{h2}+\cdots$ in the notation
of \cite{Gounaris:1997ef}) at the level of $10^{-3}$--$10^{-4}$
depending on the Higgs-boson mass. Correcting for the luminosity
adopted here ($L_{e\bar e} = 500$~fb$^{-1}$) it corresponds to
our 1-$\sigma$ uncertainty for $\alpha_{h2}$ also of the order
of $10^{-3}$--$10^{-4}$, so smaller by about two orders of
magnitude than the precision obtained here for the off-resonance
region. If, however, the Higgs boson mass is unknown, then the 
analysis presented here is applicable.
\end{itemize}

\sec{Discussions and summary}

We discussed possible new-physics search through a detailed
analysis of the
process $\gamma\gamma \to t\bar{t}\to \ell X/b X$ performed in
\cite{Grzadkowski:2003tf,Grzadkowski:2005ye} in order to find
optimal beam polarizations that minimize uncertainties in the
determination of $t\bar{t}\gamma$, $tbW$ and $\gamma\gamma H$
coupling parameters. To estimate the uncertainties we have
applied the optimal-observable method to the final
lepton/$b$-quark energy-angular distribution in $\gamma\gamma
\to t\bar{t}\to \ell X/b X$. 

Applying the optimal observable technique, we have encountered
the problem of ``unstable-solutions'' and have concluded that
there is no stable solution in the analysis trying to determine
more than
three anomalous couplings altogether. However, in contrast to
\cite{Grzadkowski:2003tf}, adopting more polarization choices
we have obtained in \cite{Grzadkowski:2005ye} some
stable solutions with three couplings.
We also found a number of two-parameter solutions,
most of which allow for the $\gaga H$- and $tbW$-couplings
determination. The expected precision of the measurement
of the Higgs coupling is of the order of $10^{-2}$ (for the scale of
new physics ${\mit\Lambda}=1$ TeV).
This shows that the $\gamma\gamma$ collider is going
to be useful for testing the Higgs sector of the SM.

Let us consider the top-quark-coupling determination in an
ideal case such that the beam polarizations could be easily
tuned and that the energy is sufficient for the on-shell
Higgs boson production. Then the best strategy would be to adjust
polarizations to construct semi-monochromatic $\gaga$ beams
such that $\sqrt{s_{\gaga}}\simeq m_H$ and on-shell Higgs
bosons are produced. This would allow for precise $\alpha_{h1,h2}$
measurement, so the virtual Higgs effects in $\gaga\to\ttbar$
would be calculable. Unfortunately,
as we have shown earlier, it is difficult to measure
$\alpha_{\gamma2}$ by looking just at $\ell X/b X$ final states
from $\gaga\to\ttbar$.
Therefore to fix $\alpha_{\gamma2}$, one should, e.g., measure the
asymmetries adopted in \cite{Choi:1995kp} to determine
the top-quark electric dipole moment which is proportional to
$\alpha_{\gamma2}$.
Then, following the analysis presented here,
one can determine $\alpha_{\gamma1}$ and $\alpha_d$. 

Finally, one must not forget that it is necessary to take into account
carefully the Standard Model contribution with radiative
corrections when trying to determine the anomalous couplings
in a fully realistic analysis. In particular this is significant
when we are interested in $C\!P$-conserving couplings since
the SM contributions there are not suppressed unlike the
$C\!P$-violating terms. On this subject, see for instance
\cite{Brandenburg:2005uu}.
 
\vspace{0.6cm}
\centerline{ACKNOWLEDGMENTS}
\vspace{0.3cm}
This work is supported in part by the State Committee for
Scientific Research (Poland) under grant 1~P03B~078~26 in
the period 2004--2006, the Grant-in-Aid for Scientific
Research No.13135219 and No.16540258 from the Japan
Society for the Promotion of Science, and the Grant-in-Aid
for Young Scientists No. 17740157 from the Ministry of
Education, Culture, Sports, Science and Technology of Japan.

\vspace*{0.5cm}

\end{document}